\documentclass[12pt,letterpaper]{article}
\usepackage{graphicx}
\usepackage{amsmath}
\usepackage{amssymb}
\usepackage{amsfonts}
\usepackage[dvipsnames]{color}
\usepackage{hyperref}

\numberwithin{equation}{section}

\newcommand{\beq}{\begin{equation}}
\newcommand{\eeq}{\end{equation}}
\newcommand{\bea}{\begin{eqnarray}}
\newcommand{\eea}{\end{eqnarray}}

\newcommand{\bC}{{\mathbb C}}

\newcommand{\bZ}{{\mathbb Z}}

\newcommand{\CN}{{\mathcal N}}
\newcommand{\CO}{{\mathcal O}}

\newcommand\tr{\mathrm{tr}}

\def\cN{\mathcal{N}}
\def\cH{\mathcal{H}}
\def\bC{\mathbb{C}}
\def\bZ{\mathbb{Z}}
\def\SU{\mathrm{SU}}
\def\U{\mathrm{U}}
\def\SO{\mathrm{SO}}
\def\vev#1{\langle #1 \rangle}
\usepackage{tikz}
\usetikzlibrary{arrows,snakes}
\tikzset{>=angle 60}
\tikzstyle{W}=[draw,circle,scale=.6]
\tikzstyle{B}=[draw,circle,fill=black,scale=.6]
\tikzstyle{H}=[draw,circle,fill=gray,scale=.6]
\tikzstyle{every picture}=[scale=.6,baseline=(current bounding box.south)]
\tikzstyle{every loop}=[]

\def\til{\widetilde}

\begin{document}

\begin{titlepage}

\begin{flushright}
SISSA 17/2012/EP\\
IPMU-12-0128 \\
UT-12-15\\
UCHEP-12-09
\end{flushright}
\vskip 1.5cm

\begin{center}
{\Large \bfseries
Quantum Higgs branches of isolated \\[4mm]
$\cN=2$ superconformal field theories
}

\vskip 1.2cm

\def\spadesuit{\alpha}
\def\heartsuit{\beta}
\def\diamondsuit{\gamma}

Philip C.~Argyres$^{\spadesuit}$, 
Kazunobu Maruyoshi$^{\heartsuit}$,
and Yuji Tachikawa$^{\diamondsuit}$

\bigskip
\bigskip

\begin{tabular}{ll}
$^{\spadesuit}$& Physics Department, University of Cincinnati,  Cincinnati OH 45221-0011, USA \\
$^{\heartsuit}$ & SISSA and INFN, Sezione di Trieste, via Bonomea 265, 34136 Trieste, Italy \\
$^{\diamondsuit}$ & Department of Physics, University of Tokyo,  Hongo, Tokyo 113-0033, Japan \\
& and Kavli IPMU,  University of Tokyo,  Kashiwa, Chiba 277-8583, Japan
\end{tabular}

\vskip 1.5cm

\textbf{Abstract}
\end{center}

\medskip
\noindent
We study the Higgs branches of the superconformal points of four-dimensional $\cN{=}2$ super Yang-Mills (SYM) which appear due to the occurrence of mutually local monopoles having appropriate charges.  
We show, for example, that the maximal superconformal point of $\SU(2n)$ SYM has a Higgs branch of the form $\bC^2/\bZ_n$.  
These Higgs branches are intrinsic to the superconformal field theory (SCFT) at the superconformal point, but do not appear in the SYM theory in which it is embedded.  
This is because the embedding is a UV extension of the SCFT in which some global symmetry acting on the Higgs branch is gauged irrelevantly.  
Higgs branches deduced from earlier direct studies of these isolated SCFTs using BPS wall-crossing or 3-d mirror symmetry agree with the ones we find here using just the Seiberg-Witten data for the SYM theories.
\bigskip
\vfill
\end{titlepage}

\section{Introduction and Summary}
\label{sec:intro}

On the Coulomb branch of a four-dimensional $\cN{=}2$ supersymmetric gauge theory, there are often points where electric and magnetic particles become simultaneously massless.  
The infrared limit of the theory at those points then becomes a superconformal theory, possibly with additional decoupled sectors \cite{Argyres:1995jj,Argyres:1995xn,Eguchi:1996vu,Eguchi:1996ds,Gaiotto:2010jf}.  
It is often the case that starting from different ultraviolet gauge theories, we end up with the same superconformal theory in the infrared limit. 
For example, the equivalence of the superconformal point of the $\SU(3)$ SYM theory and the superconformal point of the $\SU(2)$ theory with one flavor was already noted in \cite{Argyres:1995xn}. 
Additional examples of such equivalences were noted recently in \cite{Cecotti:2010fi,Bonelli:2011aa,Xie:2012hs}.
In particular, in \cite{Bonelli:2011aa} it was pointed out that 
the maximal superconformal point of the $\SO(2n)$ SYM theory is equivalent to the maximal superconformal point of the $\SU(n-1)$ theory with two flavors: their Seiberg-Witten curves and  differentials are identical. 

This equivalence presents a puzzle:  the $\SU(n-1)$ theory with two flavors has a Higgs branch of the form $\bC^2/\bZ_2$ which is rooted at the superconformal point \cite{Argyres:1996eh}, while the $\SO(2n)$ SYM has no Higgs branch at all.
Thus the maximal superconformal point of $\SO(2n)$ SYM theory should have a nontrivial Higgs branch, although the original theory in the ultraviolet does not have any hypermultiplets at all.  

The aim of this short note is to explain how a nontrivial ``quantum" Higgs branch can appear in the pure Yang-Mills gauge theory.  
We will summarize the mechanism by which a quantum Higgs branch can appear immediately below; detailed examples are discussed in later sections. 

Consider, then, an $\cN{=}2$ SYM theory with a simply-laced gauge group $G$ of rank $r$.
In the ultraviolet, we have Coulomb branch operators of the form $u_i=\tr \Phi^{d_i}$, where $d_{1,\ldots,r}$ are the degrees of the independent adjoint invariants of $G$, constructed from the scalar $\Phi$ in the vector multiplet. 
At its superconformal point, suitable redefinitions of $u_i$ have scaling dimensions $\Delta(u_i)=e_i/(h+2)$ \cite{Eguchi:1996ds}.
We distinguish three cases:
\begin{itemize}
\item $\Delta(u_i)>1$. Denote them by $v_i=u_i$, $i=1,\ldots, r_0$.
\item $\Delta(u_i)<1$. Denote them by $c_i=u_{r-i}$, $i=1,\ldots, r_0$, so that $\Delta(c_i)+\Delta(v_i)=2$.
\item $\Delta(u_i)=1$. Denote them by $m_a$, $a=1,\ldots,f=r-2r_0$.
\end{itemize}
As discussed in \cite{Argyres:1995xn}, the superconformal point has $r_0$ operators $U_i$ with vevs $v_i=\vev{U_i}$ parameterizing the Coulomb branch of the IR SCFT.  
The $c_i$ are parameters for the relevant deformations 
\begin{equation}
\delta S=\int d^4xd^4\theta\ c_i U_i,\label{c}
\end{equation}
where the integral is over the chiral $\cN{=}2$ superspace. 
The remaining $f$ $\U(1)$ multiplets on the Coulomb branch of the UV theory we denote by $M_a$, with vevs $m_a=\vev{M_a}$.  
In the IR since they have scaling dimension one at the superconformal point, they must decouple from the SCFT.  
Their vevs are mass parameters in the SCFT which explicitly break the rank-$f$ global flavor symmetry of the SCFT to its $\U(1)^f$ subgroup.

At the superconformal point, there are mutually non-local massless BPS states. 
Among them there may be $h$ mutually \emph{local} hypermultiplets, which are related to monopoles in the weakly-coupled regime. 
If it happens that $h>r_0$, then they can give rise to a nontrivial Higgs branch $\cH$. 
In these cases, it turns out that the Higgs branch $\cH$ has flavor symmetries, to which the $\U(1)$ multiplets $M_a$ couple. 
Therefore, there are $(\cN{=}2)$-invariant couplings given by an $\cN{=}1$ superpotential term 
\begin{equation}
\int d^2\theta M_a (Q\til Q)^a, \label{W}
\end{equation}
where $(Q\til Q)^a$ schematically stands for the moment map for the $a$-th $\U(1)$ flavor symmetry. 
When the multiplets $M_a$ are dynamical --- i.e., when the SCFT is embedded as an IR fixed point in the UV SYM theory --- this translates to the potential on the space $\cH$ given by  
\begin{equation}
V= (g^2)_{ab} (Q\til Q)^{ai} (Q\til Q)^{bi} + \text{terms proportional to }M_aM_b 
\end{equation}
where $i$ is the index for the $\SU(2)_R$ triplet, and $(g^2)_{ab}$ is the matrix of coupling constants of the $\U(1)$ multiplets $M_a$. 
This potential term lifts the Higgs branch when $g^2$ is nonzero.
Therefore, we have the space $\cH$ as the true Higgs branch only in the strict infrared limit where the $\U(1)$ multiplets $M_a$ are completely decoupled.

The $c_i$ parameters \eqref{c} enter the superconformal fixed point theory in the same way as do vevs of vectormultiplet fields.  
In fact, in the examples under discussion here where the UV theory is $\cN{=}2$ SYM in which all the elementary fields are in vector multiplets, the $c_i$ are literally vector multiplet vevs.  
Then a standard non-renormalization theorem \cite{Argyres:1996eh} implies that any Higgs branch which exists for non-vanishing $c_i$ will be independent of the $c_i$, and so will persist for any and all values of the $c_i$.  
Indeed, below we will demonstrate the existence of Higgs branches by deforming away from the superconformal point by turning on generic $c_i$.

It is less clear whether, in principle, there could be an additional component of the Higgs branch which occurs at the superconformal point only when the $c_i=0$.  
If the standard coupling \eqref{W} were the only coupling preserving $\cN{=}2$ supersymmetry at the superconformal point, then the potential could not depend on any parameters other than the $m_a$ and the Higgs branch at the superconformal point should persist even when we turn on the couplings $c_i$.
But it is unclear to us how to show that \eqref{W} is the only allowed coupling between a (non-Lagrangian) SCFT and $\U(1)$ multiplets.

In the rest of the paper, we study individual SYM theories in detail.
In section \ref{sec:higgs}, we study the Seiberg-Witten curves of $\SU(n+1)$ and $\SO(2n)$ SYM, and deduce the spectrum of mutually-local monopoles at the superconformal points. 
We will find that 
\begin{itemize}
\item the superconformal point of $\SU(2k)$ SYM has the Higgs branch $\bC^2/\bZ_k$;
\item the superconformal point of $\SU(2k+1)$ SYM doesn't have any Higgs branch;
\item the superconformal point of $\SO(4k+2)$ SYM has the Higgs branch $\bC^2/\bZ_2$; and
\item the superconformal point of $\SO(4k)$ SYM has a two-dimensional Higgs branch, which generically has $\SU(2)\times \U(1)$ flavor symmetry, which enhances to $\SU(3)$ only when $k=2$.
\end{itemize}

In section \ref{sec:flavor}, we study the $\SO(2n)$ SYM further. Namely, we determine the one-loop beta function of the gauge coupling of the $\U(1)$ vector multiplet $M$, which decouples at the superconformal point, directly from the Seiberg-Witten curve. 
The beta function should give the central charge of the $\SU(2)$ flavor symmetry of the superconformal point.
We will see that the central charge computed in this way indeed agrees 
with the one of the maximal superconformal point of the $\SU(n-1)$ theory with two flavors  in \cite{Gaiotto:2010jf}.

In section \ref{sec:comparisons}, we compare our findings with the properties of isolated SCFTs observed in works on the BPS spectra \cite{Cecotti:2010fi, Cecotti:2011rv, Alim:2011ae} and 3d mirror symmetry \cite{Nanopoulos:2010bv}. 
These allow two independent determinations of Higgs branches associated with large classes of isolated SCFTs.  
For the SCFTs studied here, we find they agree with each other and with the Higgs branches we compute.  
In addition, we will see that the BPS quiver method \cite{Cecotti:2010fi, Cecotti:2011rv, Alim:2011ae} predicts that the superconformal point of the $E_7$ theory will have a one-dimensional Higgs branch, but that those for $E_6$ and $E_8$ are empty. 
We have not tried to verify this using the Seiberg-Witten curves 
for the maximal superconformal points of the $E_n$ SYM theories.

One limitation of our method, mentioned above, is that it only determines those Higgs branches which persist under deformations of the SCFT by the $c_i$ parameters, but fails to rule in or out an additional Higgs branch component at $c_i=0$.  
This limitation is shared by the BPS quiver method.  
Indeed, our computations in the next section are a pedestrian approach to obtaining BPS spectra and should be subsumed by the more powerful machinery of the BPS quiver method.
The 3d mirror symmetry method, on the other hand, seems to be sensitive to data beyond the BPS spectrum.  
In particular, for our examples it predicts no additional component of the Higgs branch at $c_i=0$.

\section{Higgs branches from Seiberg-Witten curves}
\label{sec:higgs}

In this section, we consider the Higgs branches of the maximally conformal points 
of $\CN=2$ SU($n+1$) and SO($2n$) SYM theories, which we call, respectively, the $A_{n}$ and $D_{n}$ theories.
By analyzing the Seiberg-Witten curve, we explicitly show that a Higgs branch does exists exactly at these points, but disappears (is lifted) in the total SU and SO SYM theories.  
We note that the BPS spectrum of the $A_n$ superconformal points had already been studied 
in \cite{Shapere:1999xr} in a similar manner; see also a recent work \cite{Seo:2012ns}.

\subsection{$A_{n}$ theory}
\label{subsec:SU(n)}
  
First of all, we consider $\CN=2$ SU($n + 1$) SYM theory. 
The Seiberg-Witten curve is \cite{Argyres:1994xh, Klemm:1994qs}
\bea
y^{2}
=     P_{n+1}(x)^{2} - \Lambda^{2(n+1)},
\eea
where
    \bea
    P_{n+1}(x)
     =     x^{n+1} + u_{2} x^{n-1} + \ldots + u_{n} x + u_{n+1},
    \eea
and $u_{i}$ are the Coulomb moduli parameters and $\Lambda$ is the dynamical scale. 
The Seiberg-Witten differential is given by
    \bea
    \lambda_{{\rm SW}}
     =     x P'_{n+1} \frac{dx}{y},
    \eea
where $P'$ is the derivative with respect to $x$.
  
When $n \geq 2$, at a special locus of the Coulomb branch, mutually nonlocal massless particles could appear, which implies the superconformal invariance of the theory \cite{Argyres:1995jj}.
In terms of the Seiberg-Witten curve this occurs at a point where the curve strongly degenerates.
The maximally degenerate point of the above curve \cite{Argyres:1995jj, Eguchi:1996vu} is at $u_{i} = \Lambda^{n+1} \delta_{i,n+1}$, where the curve is 
    \bea
    y^{2} 
     \sim  x^{n+1} \left( x^{n+1} + 2\Lambda^{n+1} \right).
           \label{curveSU(n)AD}
    \eea
This indicates that $n+1$ branch points collide at $x = 0$ and the other $n+1$ points are at $x \sim \CO(\Lambda)$.
The relevant deformations from this point, after taking the decoupling limit $\Lambda \rightarrow \infty$ and scaling to the region near $x = 0$, is described by
    \bea
    y^{2}
     =    x^{n+1} + u_{2} x^{n-1} + \ldots + u_{n+1},
    \eea
with the Seiberg-Witten differential
    \bea
    \lambda_{{\rm SW}}
     \sim  y dx,
    \eea
where we have redefined $u_{n+1} \to \Lambda^{n+1} + u_{n+1}$.
The $n+1$ branch points at $x \sim \CO(\Lambda)$ of (\ref{curveSU(n)AD}) are effectively sent to $x = \infty$ by this procedure.

Note that the differential has a pole at $x = \infty$ of degree $(n+5)/2$ if $n$ is odd, which originates from the decoupled cuts.
Related to this fact, the properties of the $A_{n}$ theories are different depending on whether $n$ is even or odd. 
Thus, in the following, we consider them separately and show that when $n=2k-1$ the Higgs branch is $\bC^2/\bZ_k$, while there is no Higgs branch when $n=2k$. 

\subsubsection{$A_{2k-1}$ theory}
  
Let us first consider the $n=2k-1$ case ($k>1$).
The curve can be written as
    \bea
    y^{2}
     =     x^{2k} +c_{2} x^{2k-2} + \ldots + c_{k} x^{k} + c_{k+1} x^{k-1} + v_{k} x^{k-2} + \ldots  + v_{2}.
           \label{curveA2k-1}
    \eea
The scaling dimensions of the parameters are given, by demanding the dimension of the Seiberg-Witten differential is one, as
    \bea
    \Delta(c_{i})
     =     \frac{2i}{n+3}, ~~~
    \Delta(v_{i})
     =     \frac{2(n+3-i)}{n+3}, ~~~
    \Delta(c_{k+1})
     =     1,
           \label{dimensionA2k-1}
    \eea
for $i = 2, \ldots, k$.
Due to the fact that $\Delta(v_i) + \Delta(c_i) = 2$ and $\Delta(v_i) >1$, we can interpret $v_i$ and $c_i$ as the vevs of the relevant operators and their corresponding couplings as in (\ref{c}), respectively.
The Coulomb branch is $k-1$ dimensional, which is equal to the genus of the curve (\ref{curveA2k-1}).
Also, the dimension-one parameter $c_{k+1}$ is related to the mass parameter of a U(1) flavor symmetry.
Indeed, the residue at the Seiberg-Witten differential is $\frac{1}{2}c_{k+1} + f(c_i)$, 
where $f(c_i)$ is a polynomial in the $c_i$ with $i\le k$ and homogeneous of scaling dimension one.

In order to see the Higgs branch, let us set this residue to zero.
Then, for any given values of the $c_i$ parameters ($i=2, \ldots, k$), by appropriately tuning the Coulomb moduli to a special point
     \bea\label{}
     v_i
      =  v_i^\star (c_j) ,
     \eea
we can obtain the degenerate curve
     \bea\label{degc}
     y^2
      =     Q_k(x)^2,
     \eea
where $Q_k(x)$ is a polynomial of degree $k$.  
For generic (non-vanishing) $c_i$, $Q_k(x)$ will itself be non-degenerate.
Let $x_i$ ($i=1, \ldots, k$) be the distinct roots of this polynomial.
Denote by $A_i$ the $k$ degenerating cycles each of which encircles a single $x_i$.  
Only $k-1$ of these, say $A_i$ for $i =1, \ldots, k-1$, are independent homology cycles.
We take these $k-1$ $A$-cycles to define a basis for the (mutually local) charge lattice of the low energy $U(1)^{k-1}$ gauge symmetry.

Any integer linear combination of the $A_i$ cycles is a vanishing cycle of the degenerate curve (\ref{degc}).  
To determine for which of these (infinitely many) cycles there exists a BPS state in the spectrum, it is sufficient to determine the monodromy of a basis of homology cycles as one encircles $v_i^\star$ on the Coulomb branch.  This is because we are working at a generic $c_i$ where all the vanishing cycles of the degenerate curve (\ref{degc}) are mutually local, and there is a standard relation \cite{Argyres:1995jj} between these monodromies and charges of mutually local massless BPS states.

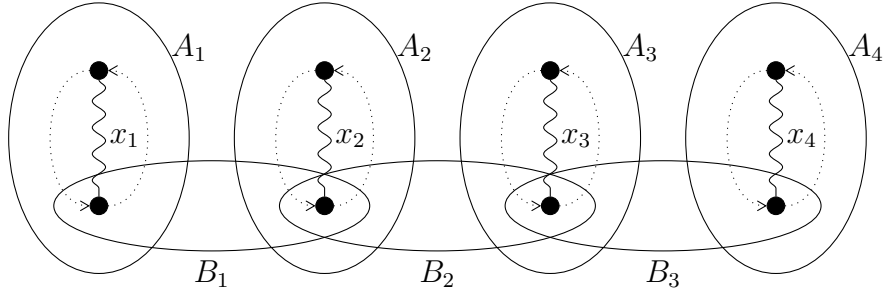
\begin{figure}
\centering
\begin{tikzpicture}
\node (A) at (0,0) {};
\node[B] (B) at (0,1.5) {};
\node[B] (C) at (0,-1.5){};
\node (P) at (5,0){};
\node[B] (Q) at (5,1.5){};
\node[B] (R) at (5,-1.5){};
\node (S) at (10,0){};
\node[B] (T) at (10,1.5){};
\node[B] (U) at (10,-1.5){};
\node (Sr) at (15,0){};
\node[B] (Tr) at (15,1.5){};
\node[B] (Ur) at (15,-1.5){};
\draw (A) node[anchor=west] {$x_1$};
\draw[->,style=dotted] (B) to[out=180,in=180] (C);
\draw[->,style=dotted] (C) to[out=0,in=0] (B);
\draw[decorate,decoration=snake] (B) -- (C);
\draw (P) node[anchor=west] {$x_2$};
\draw[->,style=dotted] (Q) to[out=180,in=180] (R);
\draw[->,style=dotted] (R) to[out=0,in=0] (Q);
\draw[decorate,decoration=snake] (Q) -- (R);
\draw (S) node[anchor=west] {$x_3$};
\draw[->,style=dotted] (T) to[out=180,in=180] (U);
\draw[->,style=dotted] (U) to[out=0,in=0] (T);
\draw[decorate,decoration=snake] (T) -- (U);
\draw (Sr) node[anchor=west] {$x_4$};
\draw[->,style=dotted] (Tr) to[out=180,in=180] (Ur);
\draw[->,style=dotted] (Ur) to[out=0,in=0] (Tr);
\draw[decorate,decoration=snake] (Tr) -- (Ur);
\draw (A) ellipse (2 and 3);
\draw (2,2) node {$A_1$};
\draw (P) ellipse (2 and 3);
\draw (7,2) node {$A_2$};
\draw (S) ellipse (2 and 3);
\draw (12,2) node {$A_3$};
\draw (Sr) ellipse (2 and 3);
\draw (17,2) node {$A_4$};
\draw (2.5,-1.5) ellipse (3.5 and 1);
\draw (2.5,-3) node {$B_1$};
\draw (7.5,-1.5) ellipse (3.5 and 1);
\draw (7.5,-3) node {$B_2$};
\draw (12.5,-1.5) ellipse (3.5 and 1);
\draw (12.5,-3) node {$B_3$};
\end{tikzpicture}
\caption{Branch points $x_i \pm b_{i} \sqrt{\delta}$, denoted by filled blobs, and cycles $A_i$, $B_i$ 
of the $A_{2k-1}$ theory; $k=4$ in the picture. The dotted arrows show the movements of the branch points under $\delta \to e^{2\pi i}\delta$. \label{fig:cutsA2k-1}}
\end{figure}

Consider a small deformation from $v_i^\star$ which deforms (\ref{degc}) to
    \bea
    y^2
     =     Q_k (x)^2 + \delta.
    \eea
The branch points now are at $x_i \pm b_i \sqrt{\delta}$, where $b_i$ are constants and we have ignored higher-order terms in $\delta$. 
Let $B_i$ ($i=1, \ldots, k-1$) be the cycles which go through $x_i$ to $x_{i+1}$ crossing the two cuts,  
as depicted in figure \ref{fig:cutsA2k-1}.
Let also the corresponding $A_{i}$ and $B_{i}$ periods be $a_i$ and $a_{Di}$ respectively.
By the monodromy transformation $\delta \rightarrow e^{2 \pi i} \delta$, we can check that $a_{Di} \rightarrow a_{Di} + a_i - a_{i+1}$.
Thus, the charges of the massless particles $q_i$ in a particular normalization of the charge lattice are obtained as in table \ref{table:A2k-1}.
Here $q_i$ is an $\CN=1$ chiral superfield (or its scalar component) and the BPS state is in an $\CN=2$ hypermultiplet consisting of $q_i$ and $\til{q}^\dagger_{i}$, where $\til{q}_i$ is another $\CN=1$ chiral superfield with the opposite gauge charges to $q_i$.  (The charges of the $\til{q}_i$ are not shown in table \ref{table:A2k-1}.)

\begin{table}
\centering
\begin{tabular}{c|ccccc}
& U(1)$_{1}$ & U(2)$_{2}$ & $\ldots$ & U(1)$_{k-2}$ & U(1)$_{k-1}$ \\
 \hline
$q_{1}$ &  1       & 0        & $\ldots$ & 0 &  0 \\
$q_{2}$ &  1       & 1        & $\ldots$ & 0 &  0 \\
$\vdots$ & $\vdots$ & $\ddots$ & $\ddots$ & $\vdots$ & $\vdots$ \\
$q_{k-1}$ &  0       & 0        & $\ldots$ & 1 &  1 \\
$q_{k}$ &  0       & 0        & $\ldots$ & 0 &  1\\
\end{tabular} 
\caption{U(1) charges of massless BPS particles of the $A_{2k-1}$ theory.\label{table:A2k-1}}
\end{table}
  
We can now see that the Higgs branch is $\bC^2/\bZ_k$ \cite{Douglas:1996sw}. 
Let us quickly recall how this works. 
Holomorphic gauge invariant operators are constructed as $M_i = q_i \til{q}_i$, $N = q_1 \til{q}_2 q_3 \til{q}_4 \cdots$, and $\til{N} = \til{q}_1 q_2 \til{q}_3 q_4 \cdots$, which satisfy the constraint $\prod_{i=1}^k M_i = N \til{N}$.
The F-term equations further give the constraints $M_i + M_{i+1} = 0$ for $i=1, \ldots, k-1$.
Therefore, the independent invariants are reduced only to $(M_1, N, \til{N}) \in \bC^3$ with the constraint
    \bea
    (- 1)^{[k/2]} M_1^k
     =     N \til{N},
    \eea
where $[s]$ is the integer part of $s$.
This is the orbifold singularity $\bC^2/\bZ_k$.
Thus, we have found a $\bC^2/\bZ_k$ Higgs branch at the special point of the Coulomb branch for arbitrary values of $c_i$.
Note that when $k=2$ the orbifold $\bC^2/\bZ_2$ has an SU(2) isometry.
Thus, the flavor symmetry is enhanced from U(1) to SU(2) in this case.
  
This Higgs branch does not exist if we go back to the original SU($2k$) description.
In particular, the U(1) flavor symmetry of the $A_{2k-1}$ theory is gauged, which can be seen from the fact that the residue at $x = \infty$ comes from decoupled cuts.
In the case of $k=2$ the U(1) subgroup of SU(2) is gauged.
Indeed, we can check that the gauge charges of $q_{1}$ and $q_{2}$ with respect to this gauged U(1) are $1$ and $-1$ respectively.

\subsubsection{$A_{2k}$ theory}
  
Let us next see that no Higgs branch exists in the $n=2k$ case.
The Seiberg-Witten curve is     
    \bea
    y^{2}
     =     x^{2k+1} +c_{2} x^{2k-1} + \ldots + c_{k+1} x^{k} + v_{k+1} x^{k-1} + \ldots  + v_{2},
           \label{curveA2k}
    \eea
and the Seiberg-Witten differential is $\lambda_{{\rm SW}} = y dx$.
Note that there is no residue at $x = \infty$ in this case.
The scaling dimensions of the parameters are given as (\ref{dimensionA2k-1}) for $i=2, \ldots, k+1$ without any dimension-one parameter.
The Coulomb branch is $k$-dimensional. 
  
We see that there is a locus in the Coulomb branch where $k$ mutually local cycles degenerate to points at generic $c_{i}$.
At this point, the curve is written as
    \bea
    y^{2}
     =     (x-x_{0}) Q_{k}(x)^{2},
    \eea
where $Q_{k}(x)$ is a polynomial of degree $k$ and $x_{0}$ is a constant.
A similar argument to the one in the $n=2k-1$ case shows that the charge assignment of massless BPS particles $q_{i}$ at this locus to be as in table \ref{table:A2k}.
    
    \begin{table}
    \centering
    \begin{tabular}{c|ccccc}
           & U(1)$_{1}$ & U(2)$_{2}$ & $\ldots$ & U(1)$_{k-1}$ & U(1)$_{k}$ \\
           \hline
     $q_{1}$ &  1       & 0        & $\ldots$ & 0 &  0 \\
     $q_{2}$ &  1       & 1        & $\ldots$ & 0 &  0 \\
     $\vdots$ & $\vdots$ & $\ddots$ & $\ddots$ & $\vdots$ &  $\vdots$ \\
     $q_{k-1}$ &  0       & 0        & $\ldots$ & 1 &  0 \\
     $q_{k}$ &  0       & 0        & $\ldots$ & 1 &  1 \\
    \end{tabular} 
    \caption{U(1) charges of massless BPS particles of the $A_{2k}$ theory    \label{table:A2k}}
    \end{table}
  
A basis of holomorphic gauge invariant operators are $M_{i} = q_{i} \til{q}_{i}$, $N$, and $\til{N}$ as before.
But the F-term equations give $M_{i} = 0$, meaning that there is no Higgs branch.
  
\subsection{$D_{n}$ theory}
\label{subsec:Dn}
  
Our second example is the $D_{n}$ theory, which is the maximally conformal point of SO($2n$) SYM theory.
We will show that this theory has SU(2) flavor symmetry in the case of odd $n$ with $\bC^2/\bZ_2$ Higgs branch, 
and SU(2)$\times$U(1) (SU(3) for $n=4$) in the case of even $n$.
For $n$ even the Higgs branch has (quaternionic) dimension two (or real dimension eight).

The Seiberg-Witten curve of SO($2n$) SYM theory is \cite{Brandhuber:1995zp, Argyres:1995fw, Hanany:1995fu}
    \bea
    \til y^{2}
     = x P_{n}(x)^{2} - \Lambda^{4(n-1)} x^{3},
           \label{curveSO(2n)}
    \eea
where 
    \bea\label{Pn}
    P_{n}(x)
     =     x^{n} + \sum_{i=1}^{n-1} s_{2i} x^{n-i} + \til s_{n}^{2},
    \eea
and the Seiberg-Witten differential is
    \bea\label{PIRf1}
    \lambda_{{\rm SW}}
     =     \left( P_n - x P'_n \right) \frac{dx}{\til y}.
    \eea
The branch points are at the $2n+1$ roots of the right hand side of (\ref{curveSO(2n)}) and at $x = \infty$.
The curve is of genus $n$.
  
The maximal degeneration of the curve occurs at $s_{2i} = \Lambda^{2(n-1)} \delta_{i,n-1}$ and $\til s_{n} = 0$ \cite{Eguchi:1996vu}, at which point the curve has the form
    \bea
    \til y^2
     \sim  \left( x^{n} + 2 \Lambda^{2(n-1)} x \right) x^{n+1}.
           \label{nearAD}
    \eea
Thus, $n+2$ branch points collapse to $x \sim 0$ and the other $n-1$ branch points are at $x \sim \CO(\Lambda^{2})$.  It is convenient to redefine $s_{2(n-1)} \to \Lambda^{2(n-1)}+s_{2(n-1)}$, so $\{s_{2j},\til s_n\}$ now measure the deviation from the singular point.
Then, by taking a decoupling limit $\Lambda \rightarrow \infty$ and scaling to the maximal singular point on the Coulomb branch, we get the curve of the $D_{n}$ maximal superconformal theory.  
Since this decoupling plus scaling procedure is less straight forward than in the $A_n$ case, we now give a few details.

Keeping only terms which control the leading positions of the zeros of $\til y^2$ near $x=0$ as all $s_j, \til s_n \to 0$, gives
\begin{align}\label{PIR1}
\til y^2 & = 2x^2 P_n + x\til s_n^4 ,
\end{align}
where $P_n$ is still given by (\ref{Pn}) even though we have shifted the definition of $s_{2(n-1)}$.  
This is equivalent to taking the $\Lambda\to\infty$ decoupling limit.
Note that at this stage the $x\til s_n^4$ term is kept even though it will later turn out to scale to zero much faster than the $x^2 \til s_n^2$ term because it nevertheless controls the position of one of the roots of $y^2$.  
Note also that (\ref{PIR1}) describes a genus $[(n+1)/2]$ curve.

Now demand a scaling symmetry to $x=\til y=s_{2j}=\til s_n=0$ in the curve.  
There is only one consistent scaling,\footnote{Thus the possibility of multiple consistent scalings, discussed in \cite{Gaiotto:2010jf}, does not arise in this case.} with $\Delta(x) = (2/n)\Delta(\til s_n) = (1/j)\Delta(s_{2j})$, which makes the $x\til s_n^4$ term in (\ref{PIR1}) irrelevant.
The curve then becomes $\til y^2 = 2x^2P_n$, which is singular at $x=0$ where one handle is always pinched, reducing the genus by one to $[(n-1)/2]$.  
This can be made explicit by defining a new coordinate $y= \til y/x$ so that the curve becomes
\begin{align}\label{PIR3}
y^2 &= 2P_n.
\end{align}
The pinched handle is no longer apparent, but will show up as a pair of poles 
in the Seiberg-Witten differential at $x=0$.
Replacing $\til y \to xy$ in (\ref{PIRf1}) gives, up to a total derivative, the differential
\begin{align}\label{PIRf3}
\lambda_\text{SW} = \frac{y}{x}\, dx
\end{align}
on the part of the curve in the vicinity of the singularity.
Demanding that $\Delta(\lambda_\text{SW})=1$ then gives 
\begin{align}\label{dnsd}
\Delta(s_{2j})=\frac{2j}{n} 
\qquad\text{and}\qquad 
\Delta(\til s_n)=1.
\end{align}
$\lambda_\text{SW}$ has a pole at $x=0$ with residue $\sim \til s_n$, which identifies $\til s_n$ as a mass parameter.  
The $s_{2j}$ with $2j >n$ have dimensions greater than one, so are identified with Coulomb branch vevs, while those with $2j<n$ have dimensions less than one, so are relevant couplings like the $c_i$ in (\ref{c}).

Note that for $n$ even, $s_n$ has dimension one, 
and $\lambda_\text{SW}$ also has a pole at $x=\infty$ with residue $\sim s'_n := s_n + f(s_{2j})$ where $f$ is a polynomial in the $s_{2j}$ with $2j < n$ which is homogeneous of scaling dimension one.  
This identifies $s'_n$ as a second mass parameter in the theory for $n$ even.
As in the $A_n$ theory, let us consider the odd and even $n$ cases separately.

\subsubsection{$D_{2k+1}$ theory}
\label{subsec:SO(4n+2)}
  
We first consider the $n=2k+1$ case.
The curve (\ref{PIRf3}) is (after rescaling $y$ by $\sqrt2$ and renaming the $\{s_{2j},\til s_n\}$ parameters)
    \bea\label{curveADSO(4k+2)}
    y^{2}
     =     x^{2k+1} + c_{1} x^{2k} + \ldots + c_{k} x^{k+1}
         + v_{k} x^{k} + \ldots + v_1 x + m^{2}.
    \eea
As in the $A_n$ case, the $v_{i}$ are the Coulomb moduli and $c_{i}$ are the associated deformation parameters.
Their scaling dimensions (\ref{dnsd}) are thus $\Delta(c_i) = (2i/n)$,  $\Delta(v_i) = 2(n-i)/n$, and $\Delta(m) = 1$, for $i=1, \ldots, k$.
Note that this curve is the same as that of (the relevant deformation from) the maximal superconformal point of the SU($2k$) theory with two flavors \cite{Bonelli:2011aa}.
So, it might be natural to expect that the flavor symmetry is SU(2).
Let us check this here.

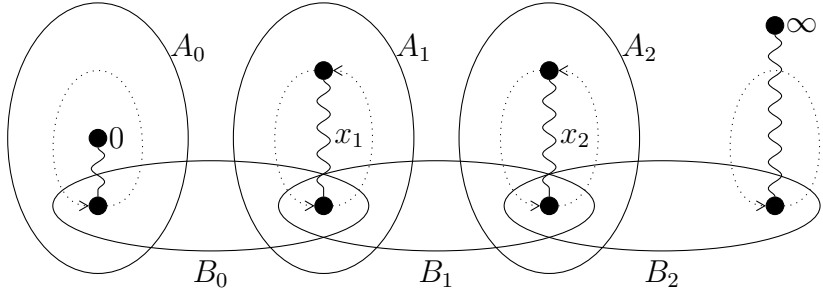
\begin{figure}
\centering
\begin{tikzpicture}
\node[B] (A) at (0,0) {};
\node (B) at (0,1.5) {};
\node[B] (C) at (0,-1.5){};
\node[B] (Q) at (5,1.5){};
\node[B] (R) at (5,-1.5){};
\node[B] (T) at (10,1.5){};
\node[B] (U) at (10,-1.5){};
\node (X) at (15,1.5){};
\node[B] (Y) at (15,-1.5){};
\node[B] (Z) at (15,2.5){};
\draw[decorate,decoration=snake] (A) -- (C);
\draw[decorate,decoration=snake] (Q) -- (R);
\draw[decorate,decoration=snake] (T) -- (U);
\draw[decorate,decoration=snake] (Y) -- (Z);
\draw (A) node[anchor=west] {$0$};
\draw[->,style=dotted] (C) to[out=0,in=0] (B.center) to[out=180,in=180] (C) ;
\draw (P) node[anchor=west] {$x_1$};
\draw[->,style=dotted] (Q) to[out=180,in=180] (R);
\draw[->,style=dotted] (R) to[out=0,in=0] (Q);
\draw (S) node[anchor=west] {$x_2$};
\draw[->,style=dotted] (T) to[out=180,in=180] (U);
\draw[->,style=dotted] (U) to[out=0,in=0] (T);
\draw (Z) node[anchor=west] {$\infty$};
\draw[->,style=dotted] (Y) to[out=0,in=0] (X.center) to[out=180,in=180] (Y) ;
\draw (A) ellipse (2 and 3);
\draw (2,2) node {$A_0$};
\draw (P) ellipse (2 and 3);
\draw (7,2) node {$A_1$};
\draw (S) ellipse (2 and 3);
\draw (12,2) node {$A_2$};
\draw (2.5,-1.5) ellipse (3.5 and 1);
\draw (2.5,-3) node {$B_0$};
\draw (7.5,-1.5) ellipse (3.5 and 1);
\draw (7.5,-3) node {$B_1$};
\draw (12.5,-1.5) ellipse (3.5 and 1);
\draw (12.5,-3) node {$B_2$};
\end{tikzpicture}
\caption{Branch points and cycles of the $D_{2k+1}$ theory when $m=0$; $k=3$ in the picture. 
The dotted arrows show the movements of the branch points under $\delta \to e^{2\pi i}\delta$. \label{fig:cutsD2k+1}}
\end{figure}
  
As in the $A_{n}$ theory, we set the mass parameter to zero and choose the special locus of $v_i=v_i^\star(c_j)$ where $k$ non-intersecting cycles degenerate.
Note that this locus satisfies $v_1^\star=0$. 
We can see that at this locus there are $k+1$ mutually local massless BPS particles by examining a monodromy around the locus in the Coulomb branch.
Indeed, by setting $v_i=v_i^\star$ for arbitrary fixed $c_{i}$, the Seiberg-Witten curve degenerates to
    \bea
    y^{2}
     =     x^{2} (x - x_{k}) Q_{k-1}(x)^{2},
    \eea
where $Q_{k-1}(x)$ is a degree-$(k-1)$ polynomial whose roots we denote as $x_{i}$ ($i=1, \ldots, k-1$) and $x_{k}$ is a constant depending on the $c_{i}$. 
A small deformation from this locus (still fixing $m=0$) is 
    \bea
    y^{2}
     =     x \left( x(x-x_{k}) Q_{k-1}(x)^{2} - \delta \right).
    \eea
The branch points  are at $x = 0, \delta$, $x_{i} \pm b_{i}\sqrt{\delta}$ and $x_{k}+ b_{k} \delta$ where $b_{i}$ are some constants.
Let $A_{0}$ and $A_{i}$ ($i= 1, \ldots, k-1$) be the cycles around $x = 0, \delta$ and $x=x_{i} \pm b_{i} \sqrt{\delta}$, respectively.
Let also $B_{0}$ and $B_{i}$ ($i= 1, \ldots, k-1$) be the cycles from $0$ to $x_{1}$ and from $x_{i}$ to $x_{i+1}$ passing through two cuts, respectively; see figure \ref{fig:cutsD2k+1}.

The monodromy under $\delta \rightarrow e^{2 \pi i} \delta$ is  $a_{D0} \rightarrow a_{D0} + 2a_{0} - a_{1}$, $a_{Di} \rightarrow a_{Di} + a_{i} - a_{i+1}$ and $a_{Dk-1} \rightarrow a_{Dk-1} + a_{k-1}$, while the $a_{i}$ periods stay fixed.
Therefore the charges of the $k+1$ massless BPS particles are assigned as in table \ref{table:D2k+1}.
Note that the coefficient 2 in $a_{D0} \rightarrow a_{D0} + 2a_{0} - \cdots$ signifies the existence of two hypermultiplets with charge 1, because in general we have $a_D\to a_D + k a$ where $k=\sum_s q_s{}^2$ where $q_s$ are the charges of the hypermultiplets charged under the $\U(1)$ corresponding to $a$. 
 
\begin{table}
\centering
\begin{tabular}{c|ccccc}
& U(1)$_{1}$ & U(2)$_{2}$ & $\ldots$ & U(1)$_{k-1}$ & U(1)$_{k}$ \\
\hline
$q_{1}$ &  1       & 0        & $\ldots$ & 0 &  0 \\
$q_{2}$ &  1       & 0        & $\ldots$ & 0 &  0 \\
$q_{3}$ &  1       & 1        & $\ldots$ & 0 &  0 \\
$\vdots$ & $\vdots$ & $\ddots$ & $\ddots$ & $\ddots$ &  $\vdots$ \\
$q_{k+1}$ &  0       & 0        & $\ldots$ & 1 &  1 \\
\end{tabular} 
\caption{U(1) charges of massless particles of the $D_{2k+1}$ theory.\label{table:D2k+1}}
\end{table}
  
This charge assignment shows that the Higgs branch is simply $\bC^{2}/\bZ_{2}$.  
Indeed, the gauge invariant operators constructed from $q_{i}$ are $M_{i} = q_{i} \til{q}_{i}$ ($i= 1, \ldots, k+1$) and $N = q_{1} \til{q}_{2}$ and $\til{N} = q_{2} \til{q}_{1}$, with one constraint $M_{1} M_{2} = N \til{N}$.
The F-term equations set $M_{i} = 0$ for $i=3, \ldots, k+1$ and $M_{1} = - M_{2}$.
Thus, $M_{1}^{2} + N \til{N} =0$, implying $\bC^{2}/\bZ_{2}$.
The flavor symmetry is SU(2) under which the BPS particles $q_{1}$ and $q_{2}$ transform as a doublet.

Next, in order to see how this SU(2) flavor symmetry is broken in the total SO($4k+2$) SYM theory, 
we analyze the spectrum of light BPS particles with a small mass parameter $m$.
While it could be possible to do this for general $k$, we consider only $k=1$ for illustration.
In this case, the curve is $y^{2} = x^{3} + c_{1} x^{2} + v_{1} x + m^{2}$ and its discriminant is 
    \bea
    \Delta
     =   - 4 \left( v^{3}_{1} - \frac{c^{2}_{1}}{4} v^{2}_{1} - \frac{9c_{1} m^{2}}{2} v_{1}
         + c_{1}^{3} m^{2} + \frac{27}{4} m^{4} \right).
    \eea
The roots of the discriminant are $v_{1} = w_{0}$ and $w_{\pm}$ where
    \bea
    w_{0}
     =       \frac{c_{1}^{2}}{4} + \frac{2}{c_{1}} m^{2} + \CO(m^{3}), ~~
    w_{\pm}
     =     \pm 2 \sqrt{c_{1}} m - \frac{1}{c_{1}} m^{2} + \CO(m^{3}).
    \eea
The last two roots are related by the sign flip of $m$.
Let us focus on the point $v_{1} = w_{+}$ where the three roots of the right-hand side of the curve are at $x \sim \CO(m)$, $\CO(m)$ and $\CO(1)$.
The $A_{0}$ cycle collapses at this value of $v_{1}$ and this indicates a massless BPS particle with central charge $- m - \hat{a}$, where $\hat{a}$ is the period of the cycle along the other cut (from $x\sim\CO(m)$ to $\infty$).
It is easy to see that the point $v_{1} = w_{-}$ corresponds to the massless particle with central charge $m - \hat{a}$ since this is obtained by the sign flip of the above case.
(When $m=0$ these two BPS particles become the doublet of the SU(2) flavor symmetry.)
Thus, these two BPS particles have charge $+1$ and $-1$ under the U(1) subgroup of SU(2).
  
As we saw above, the pole of $\lambda_{{\rm SW}}$ at $x=0$ comes from a degenerating cut of the original Seiberg-Witten curve (\ref{PIR1}) or (\ref{curveSO(2n)}).
Thus, the U(1) subgroup of SU(2) at the maximally conformal point is gauged in the total SO(6) theory. 
This argument can be applied to the general SO($4k+2$) case (and also to the SO($4k$) case which will be discussed in the next subsection): the U(1) subgroup of SU(2) is gauged and there is no Higgs branch.

\subsubsection{$D_{2k}$ theory}
\label{subsubsec:SO(4n)}
  
We now turn to the $D_{2k}$ theory.
The curve is
    \bea
    y^{2}
     =     x^{2k} + c_{1} x^{2k-1} + \ldots + c_{k-1} x^{k+1} + c_{k} x^{k}
         + v_{k-1} x^{k-1} + \ldots + v_1 x + m^{2}.
           \label{curveADSO(4k)}
    \eea
The Seiberg-Witten differential is again $\lambda_{{\rm SW}} = (y/x) dx$ with $m$ the residue of the pole at $x=0$.  
The enhancement of the U(1) flavor symmetry associated to $m$ to an SU(2) flavor symmetry follows by repeating the argument in the previous subsection.
  
The new feature here is the existence of the additional pole at $x = \infty$.
Its residue can be written in terms of $c_{k}$ plus a polynomial in the other $c_{i<k}$ parameters, 
and has scaling dimension one.  
This means that, up to a shift, $c_k$ is a mass parameter with an associated U(1) flavor symmetry.
Thus the total flavor symmetry of the $D_{2k}$ superconformal point  is at least SU(2)$\times$U(1).

By setting $m=0$, we can find the locus of the Coulomb moduli where the Seiberg-Witten curve degenerates to
    \bea
    y^2
     =     x^{2} Q_{k-1}(x)^2,
    \eea
where $Q_{k-1}(x)$ is again a degree-$(k-1)$ polynomial.
This represents the appearance of $k+1$ mutually local massless particles $q_{i}$.
This can be seen in the same way as in the $n=2k+1$ case, and their charge assignments are given in table \ref{table:D2k}.   

\begin{table}
\centering
\begin{tabular}{c|ccccc}
             & U(1)$_{1}$ & U(2)$_{2}$ & $\ldots$ & U(1)$_{k-2}$ & U(1)$_{k-1}$ \\
\hline
$q_{1}$ &  1       & 0        & $\ldots$ & 0 &  0 \\
$q_{2}$ &  1       & 0        & $\ldots$ & 0 &  0 \\
$q_{3}$ &  1       & 1        & $\ldots$ & 0 &  0 \\
$\vdots$ & $\vdots$ & $\ddots$ & $\ddots$ & $\ddots$ &  0 \\
$q_{k}$ &  0       & 0        & $\ldots$ & 1 &  1 \\
$q_{k+1}$ &  0       & 0        & $\ldots$ & 0 &  1
\end{tabular} 
\caption{U(1) charges of massless particles of the $D_{2k}$ theory.\label{table:D2k}}
\end{table}
  
Note that the flavor symmetry is enhanced to SU(3) in the $k=2$ case, that is the $D_{4}$ theory.
In this case, the Weyl group of the $\SU(3)$ flavor symmetry can be identified with the $S_3$ outer automorphism group of the original $D_4=\SO(8)$ theory. 
In the other theories, the flavor symmetry is SU(2)$\times$U(1).
The Higgs branch is two-(quaternionic-)dimensional (or, 8-real-dimensional).
In the total SO($4k$) theory, the $\U(1)^2 \subset \SU(2)\times\U(1)$ (or SU(3)) is gauged and there is no Higgs branch.

\section{Flavor central charge}
\label{sec:flavor}
  We will consider in this section the central charge of the SU(2) flavor symmetry of the $D_{n}$ theory 
  by embedding it into the SO($2n$) SYM theory
  where the U(1) subgroup of the SU(2) is gauged and by analyzing the Seiberg-Witten curve.
  
  Let us suppose that we have a flavor symmetry $H$.
  The flavor central charge is defined by the OPE of the currents $J_{\mu}^{a}$ as 
    \bea
    J^{a}_{\mu}(x) J^{b}_{\nu}(0)
     =     \frac{3k_{H}}{4 \pi^{4}} \delta^{ab} \frac{x^{2} g_{\mu \nu} - 2 x_{\mu} x_{\nu}}{x^{8}} + \ldots.
    \eea
  We normalize the flavor central charge such that $n$ free chiral multiplets give $k_{\U(n)} = 1$.
  If there is a weakly gauged group $G$ which is a subgroup of the flavor symmetry $H$,
  the central charge is given by
    \bea
    k_{G \subset H}
     =     I_{G \hookrightarrow H} k_{H},
    \eea
  where $I_{G \hookrightarrow H}$ is the embedding index, see e.g.~\cite{Argyres:2007cn}.
  Then, the one-loop beta function of the gauge group $G$ is given in terms of the central charge:
    \bea
    b_{0}
     =     2 T_{2}({\bf adj}) - \frac{k_{G \subset H}}{2}.
    \eea
  where we defined $b_{0} = \frac{\partial}{\partial \ln Q} ( 8 \pi^{2}/g^{2}(Q) )$.
  When $G=\U(1)$ and $H=\SU(2)$ which is the case we will see below,
  we have $k_{\U(1) \subset \SU(2)} = I_{\U(1) \hookrightarrow \SU(2)} k_{{\rm SU(2)}} = k_{{\rm SU(2)}}$.
  Thus, the beta function coefficient of the U(1) gauge group is $b_{0} = - \frac{k_{{\rm SU(2)}}}{2}$.
  In other words, the effective coupling constant of the U(1) should be written as
    \bea
    \tau
     =     \frac{b_{0}}{2 \pi i } \ln \Lambda
     =   - \frac{k_{{\rm SU(2)}}}{4 \pi i} \ln \Lambda.
           \label{tau}
    \eea

\begin{figure}
\centering
\begin{tikzpicture}
\node[B] (A) at (0,0) {};
\node (B) at (0,1.5) {};
\node[B] (C) at (0,-1.5){};
\node[B] (Q) at (5,1.5){};
\node[B] (R) at (5,-1.5){};
\node[B] (F) at (10,1.5){};
\node[B] (G) at (10,-1.5){};
\node (L) at (15,0){};
\node[B] (T) at (15,1.5){};
\node[B] (U) at (15,-1.5){};
\node (X) at (20,1.5){};
\node[B] (Y) at (20,-1.5){};
\node[B] (Z) at (20,2.5){};
\draw[decorate,decoration=snake] (A) -- (C);
\draw[decorate,decoration=snake] (Q) -- (R);
\draw[decorate,decoration=snake] (F) -- (G);
\draw[decorate,decoration=snake] (T) -- (U);
\draw[decorate,decoration=snake] (Y) -- (Z);
\draw (A) node[anchor=west] {$0$};
\draw (P) node[anchor=west] {$x_0$};
\draw (S) node[anchor=west] {$x_1$};
\draw (L) node[anchor=west] {$x_2$};
\draw (Z) node[anchor=west] {$\infty$};
\draw (A) ellipse (1.5 and 3);
\draw (2,2) node {$\hat{A}$};
\draw (P) ellipse (1.5 and 3);
\draw (7,2) node {$A_0$};
\draw (S) ellipse (1.5 and 3);
\draw (12,2) node {$A_1$};
\draw (L) ellipse (1.5 and 3);
\draw (17,2) node {$A_2$};
\draw (2.5,-1.5) ellipse (3.5 and 1);
\draw (2.5,-3) node {$\hat{B}$};
\draw (7.5,-1.5) ellipse (3.5 and 1);
\draw (7.5,-3) node {$B_0$};
\draw (12.5,-1.5) ellipse (3.5 and 1);
\draw (12.5,-3) node {$B_1$};
\draw (17.5,-1.5) ellipse (3.5 and 1);
\draw (17.5,-3) node {$B_2$};
\end{tikzpicture}
\caption{Branch points and cycles of the curve (\ref{curvealmostdecouple}) for $n=7$, for $m\ne 0$. The $\hat{A}$ cycle  degenerates in the limit $\Lambda \rightarrow \infty$ corresponding to that the U(1) gauge group is decoupled. \label{fig:cutsDn}}
\end{figure}
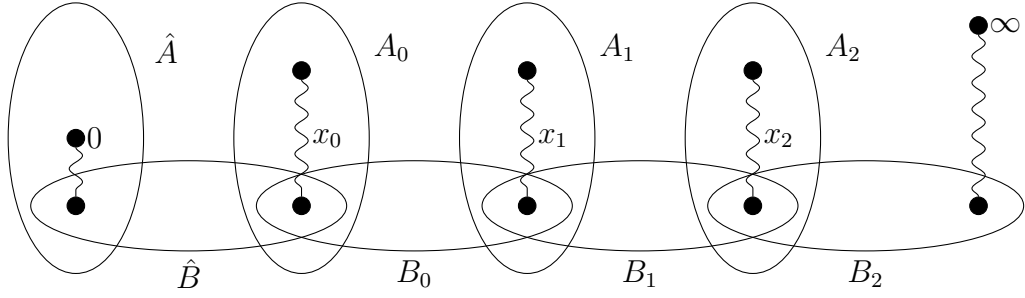

  Now let us consider the $D_{n}$ theory which has $\SU(2)$($\times \U(1)$) flavor symmetry (when $n=2k$).
  As we discussed in detail in subsection \ref{subsec:Dn}, the U(1) subgroup of the SU(2) is gauged
  in the IR effective theory of the SO($2n$) SYM theory where the $D_{n}$ theory is embedded.
  This U(1) gauge symmetry corresponds to the degenerating cut (or the pinched handle) 
  in the decoupling limit $\Lambda \rightarrow \infty$.
  In order to see this more clearly, let us go back to the curve (\ref{PIR1}) and keep the dynamical scale dependence explicit:
    \bea
    \tilde{y}^{2}
     \sim  x \left( 2 \Lambda^{2(n-1)} P_{n} + \tilde{s}_{n}^{4} \right)
     \sim  2 \Lambda^{2(n-1)}  x \left(x + \frac{\tilde{s}_{n}^{2}}{2 \Lambda^{2(n-1)}} \right) P_{n}.
    \eea
  By scaling $x$ and the parameters appropriately, we get
    \bea
    \tilde{y}^{2}
     =     x \left( x + \frac{m^{2}\Lambda^{-2(n-1)/n}}{2} \right) P_{n},
           \label{curvealmostdecouple}
    \eea
  with the differential $\lambda_{{\rm SW}} = \frac{\tilde{y}}{x^{2}} dx$,
  where $P_{n}$ is the RHS of (\ref{curveADSO(4k+2)}) for odd $n$ or (\ref{curveADSO(4k)}) for even $n$.
  The degenerating cut is from $x = 0$ to $x = - m^{2} \Lambda^{-2(n-1)/n}/2$.
  Denote by $\hat{A}$ and $\hat{B}$ the cycles encircling this cut 
  and from $x = - m^{2} \Lambda^{-2(n-1)/n}/2$ to $x = x_{0}$ going through two cuts, as depicted in figure \ref{fig:cutsDn}.
  The period integrals in the limit where $\Lambda \rightarrow \infty$ can be evaluated as follows:
    \bea
    \oint_{\hat{A}} \lambda_{{\rm SW}}
    &=&    2 \pi i m,
           \nonumber \\
    \oint_{\hat{B}} \lambda_{{\rm SW}}
    &\sim& \int^{- m^{2} \Lambda^{-2(n-1)/n}/2} m \frac{dx}{x}
     \sim- \frac{2(n-1)}{n} m \ln \Lambda.
           \label{AandB}
    \eea
  The coupling constant of the decoupling $\U(1)$ is then given by 
    \bea
    \hat{\tau}
     =     
         - \frac{1}{4 \pi i} \frac{4(n-1)}{n} \ln \Lambda + \ldots.
    \eea
  It follows from this and (\ref{tau}) that the central charge of the flavor symmetry is
    \bea
    k_{{\rm SU(2)}}
     =     \frac{4(n-1)}{n}.
    \eea
  which agrees with the result in \cite{Gaiotto:2010jf}.

\section{Comparisons to other recent results}
\label{sec:comparisons}

The properties of the Higgs branch of the superconformal points of $\cN{=}2$ SYM theories can also be determined from more modern methods. 

\subsection{Using BPS quivers}

\begin{figure}
\[
\begin{array}{clclcl}
A_3 & \begin{tikzpicture}
\node[W] (1) at (0,0) {};
\node[B] (2) at (1,0) {};
\node[W] (3) at (2,0) {};
\path (1) edge[->] (2);
\path (2) edge[<-] (3);
\end{tikzpicture} &
D_4 & \begin{tikzpicture}[baseline=(4).north]
\node[W] (1) at (0,0) {};
\node[B] (2) at (1,0) {};
\node[W] (3) at (2,.5) {};
\node[W] (4) at (2,-.5) {};
\path (1) edge[->] (2);
\path (3) edge[->] (2);
\path (4) edge[->] (2);
\end{tikzpicture} &
E_6 & \begin{tikzpicture}
\node[B] (1) at (0,0) {};
\node[W] (2) at (1,0) {};
\node[B] (3) at (2,0) {};
\node[W] (A) at (2,1) {};
\node[W] (4) at (3,0) {};
\node[B] (5) at (4,0) {};
\path (1) edge[<-] (2);
\path (2) edge[->] (3);
\path (3) edge[<-] (4);
\path (3) edge[<-] (A);
\path (4) edge[->] (5);
\end{tikzpicture} \\
A_4 & \begin{tikzpicture}
\node[W] (1) at (0,0) {};
\node[B] (2) at (1,0) {};
\node[W] (3) at (2,0) {};
\node[B] (4) at (3,0) {};
\path (1) edge[->] (2);
\path (2) edge[<-] (3);
\path (3) edge[->] (4);
\end{tikzpicture} &
D_5 & \begin{tikzpicture}[baseline=(4).north]
\node[B] (0) at (-1,0) {};
\node[W] (1) at (0,0) {};
\node[B] (2) at (1,0) {};
\node[W] (3) at (2,.5) {};
\node[W] (4) at (2,-.5) {};
\path (0) edge[<-] (1);
\path (1) edge[->] (2);
\path (3) edge[->] (2);
\path (4) edge[->] (2);
\end{tikzpicture} &
E_7 & \begin{tikzpicture}
\node[W] (0) at (-1,0) {};
\node[B] (1) at (0,0) {};
\node[W] (2) at (1,0) {};
\node[B] (3) at (2,0) {};
\node[W] (A) at (2,1) {};
\node[W] (4) at (3,0) {};
\node[B] (5) at (4,0) {};
\path (1) edge[<-] (0);
\path (1) edge[<-] (2);
\path (2) edge[->] (3);
\path (3) edge[<-] (4);
\path (3) edge[<-] (A);
\path (4) edge[->] (5);
\end{tikzpicture} \\
A_5 & \begin{tikzpicture}
\node[W] (1) at (0,0) {};
\node[B] (2) at (1,0) {};
\node[W] (3) at (2,0) {};
\node[B] (4) at (3,0) {};
\node[W] (5) at (4,0) {};
\path (1) edge[->] (2);
\path (2) edge[<-] (3);
\path (3) edge[->] (4);
\path (4) edge[<-] (5);
\end{tikzpicture}&
D_6 & \begin{tikzpicture}[baseline=(4).north]
\node[W] (-1) at (-2,0) {};
\node[B] (0) at (-1,0) {};
\node[W] (1) at (0,0) {};
\node[B] (2) at (1,0) {};
\node[W] (3) at (2,.5) {};
\node[W] (4) at (2,-.5) {};
\path (-1) edge[->] (0);
\path (0) edge[<-] (1);
\path (1) edge[->] (2);
\path (3) edge[->] (2);
\path (4) edge[->] (2);
\end{tikzpicture}&
E_8 & \begin{tikzpicture}
\node[B] (-1) at (-2,0) {};
\node[W] (0) at (-1,0) {};
\node[B] (1) at (0,0) {};
\node[W] (2) at (1,0) {};
\node[B] (3) at (2,0) {};
\node[W] (A) at (2,1) {};
\node[W] (4) at (3,0) {};
\node[B] (5) at (4,0) {};
\path (-1) edge[<-] (0);
\path (1) edge[<-] (0);
\path (1) edge[<-] (2);
\path (2) edge[->] (3);
\path (3) edge[<-] (4);
\path (3) edge[<-] (A);
\path (4) edge[->] (5);
\end{tikzpicture} 
\end{array}
\]
\caption{BPS quivers of the superconformal points of the $A$-$D$-$E$ SYM theories.\label{fig:BPSquiver}}
\end{figure}
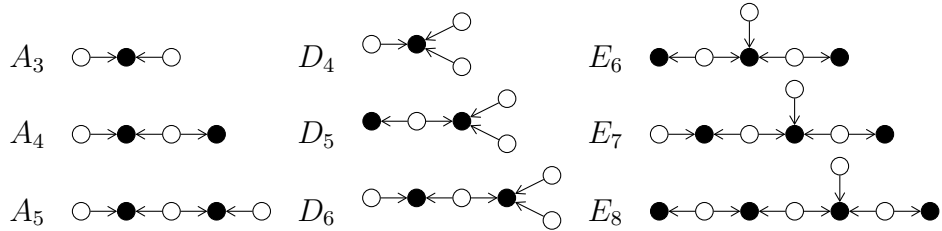

In \cite{Cecotti:2010fi}, it was stated that the BPS spectrum in a certain chamber on the Coulomb branch of the superconformal point of the SYM theory with simply-laced gauge group $G$ is given by a BPS quiver of the form of the Dynkin diagram of type $G$; see figure \ref{fig:BPSquiver}.
Namely, each node stands for a hypermultiplet, and two nodes are connected by $n=|\langle q_1,q_2\rangle|$ arrows, where $q_{1,2}$ are electromagnetic charges of two particles, and $\langle\cdot,\cdot\rangle$ is the Dirac quantization pairing; the arrow is oriented by the sign of $\langle q_1,q_2\rangle$.
For $G=\SU(N)$ this was already known in \cite{Shapere:1999xr}; our analysis in section \ref{subsec:Dn} can be thought of as an  elementary confirmation of this statement for $G=\SO(2n)$.

Any simply-laced Dynkin diagram is bipartite; let us color the nodes accordingly with white and black. 
Then the particles corresponding to the white nodes are mutually local, and similarly for the particles corresponding to the black nodes. 
Let us say that the number of black nodes is not larger than the number of white nodes. 
Let us then take the basis of U$(1)$ charges such that the $i$-th black node has the magnetic charge $\delta_{ij}$ under the $j$-th $\U(1)$ and no electric charges.
Then the $i$-th white node has the electric charge $+1$ under $j$-th $\U(1)$ charge if the $i$-th white node and the $j$-th black node are connected, and neutral under the $j$-th $\U(1)$ otherwise. 
This procedure easily reproduces the tables of charges in tables \ref{table:A2k-1}, \ref{table:A2k}, \ref{table:D2k+1}, and \ref{table:D2k}.
This also makes it manifest that the enhancement of the flavor symmetry to $\SU(3)$ is only possible when $G=D_4$.

The only thing that our analysis in section \ref{sec:higgs} adds to this picture is the direct construction from the Seiberg-Witten curve that of a locus on the Coulomb branch where the BPS states corresponding to the white modes are massless while those corresponding to the black nodes are massive.  To do this it was crucial that we could deform away from the superconformal point by turning on the relevant $c_i$ couplings (\ref{c}).

Assuming that the BPS spectrum of the superconformal points of  $E_n$ SYM theory, first studied in \cite{Eguchi:1996ds}, 
is given by the Dynkin diagram, we can immediately find the table of $\U(1)$ charges of mutually local particles. 
We do not have Higgs branches for $E_6$ and $E_8$, while we have a one-dimensional Higgs branch for $E_7$. 
This also is in accord with the spectrum of scaling dimensions of these theories, because there is no dimension-one operator when $G=E_{6,8}$, while there is one dimension-one operator when $G=E_7$.

\subsection{Using the 3d mirror}

\begin{figure}
\[
\begin{array}{cccccccccccc}
A_3 & \begin{tikzpicture}
\node[H] (1) at (0,0) {};
\node[H] (2) at (1,0) {};
\path (1) edge[in=150,out=30] (2);
\path (1) edge[in=-150,out=-30] (2);
\end{tikzpicture} &
A_5 & \begin{tikzpicture}
\node[H] (1) at (0,0) {};
\node[H] (2) at (1,0) {};
\path (1) edge[in=150,out=30] (2);
\path (1) edge (2);
\path (1) edge[in=-150,out=-30] (2);
\end{tikzpicture}&
A_7 & \begin{tikzpicture}
\node[H] (1) at (0,0) {};
\node[H] (2) at (1,0) {};
\path (1) edge[in=150,out=30] (2);
\path (1) edge[in=170,out=10] (2);
\path (1) edge[in=-170,out=-10] (2);
\path (1) edge[in=-150,out=-30] (2);
\end{tikzpicture}&
D_4 & \begin{tikzpicture}
\node[H] (1) at (0,0) {};
\node[H] (2) at (1,0) {};
\node[H] (3) at (.5,-1) {};
\path (1) edge (2);
\path (3) edge (1);
\path (3) edge (2);
\end{tikzpicture}&
D_6 & \begin{tikzpicture}
\node[H] (1) at (0,0) {};
\node[H] (2) at (1,0) {};
\node[H] (3) at (.5,-1) {};
\path (1) edge[in=150,out=30] (2);
\path (1) edge[in=-150,out=-30] (2);
\path (3) edge (1);
\path (3) edge (2);
\end{tikzpicture}&
D_8 & \begin{tikzpicture}
\node[H] (1) at (0,0) {};
\node[H] (2) at (1,0) {};
\node[H] (3) at (.5,-1) {};
\path (1) edge[in=150,out=30] (2);
\path (1) edge (2);
\path (1) edge[in=-150,out=-30] (2);
\path (3) edge (1);
\path (3) edge (2);
\end{tikzpicture}
\end{array}
\]
\caption{3d mirrors of the $A_{2k-1}$ and $D_{2k}$ theories. \label{fig:mirror}}
\end{figure}
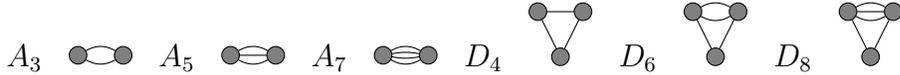

In the mathematical works \cite{Boalch1,Boalch2}, it was shown that the 3d \emph{Coulomb} branch of a wild quiver gauge theory (in the sense of \cite{Bonelli:2011aa}) compactified on $S^1$ of radius $R$ is given, in the limit $R\to 0$, by the \emph{Higgs} branch of another non-wild quiver gauge theory.  In section 6 of \cite{Nanopoulos:2010bv}, this non-wild quiver gauge theory was interpreted as the 3d mirror of the $S^1$ compactification of the wild quiver gauge theory. Then, some properties of the Higgs branches of the original theory are visible using the Coulomb branch of the 3d mirror, see figure \ref{fig:mirror}. 
In the figure, a gray node stands for a $\U(1)$ gauge group, and a line between two nodes corresponds to a bifundamental hypermultiplet; the overall $\U(1)$ is decoupled and to be removed. 
We see that there is a one-dimensional Higgs branch when $G=A_{2k-1}$, and a two-dimensional Higgs branch when $G=D_{2k}$. 

The flavor symmetry enhancement on the Coulomb branch side of a quiver was studied in \cite{Gaiotto:2008ak}; the rule of thumb is that the set of nodes corresponding to a $\U(N)$ group coupled to precisely $2n$ fundamentals, together with the edges among them, forms a finite or affine Dynkin diagram $\Gamma$.
Then the flavor symmetry enhancement is given by the type of $\Gamma$.
In our case all the nodes correspond to $\U(1)$ groups, so $N=1$.
We then easily see that for $G=A_3$ we have $\SU(2)$ flavor symmetry, while for other $G=A_{2k-1}$ we just have $\U(1)$ flavor symmetry. 
We also see that $G=D_4$ we have $\SU(3)$ flavor symmetry, while for other $G=D_{2k}$ we just have $\SU(2) \times \U(1)$ flavor symmetry.
Note that this method explains only the flavor symmetry enhancement, and we have supplied a U(1) to the generic $A_{2k-1}$ and $D_{2k}$ cases because the ranks of the flavor symmetries are always one and two respectively.



\section*{Acknowledgments}
The authors would like to thank 
G.~Bonelli, S.~Cecotti, T.~Eguchi, A.~Shapere, A.~Tanzini, and D.~Xie
for useful comments. 
PCA is partially supported by DOE grant FG02-84-ER40153.
KM is partially supported by the INFN project TV12. 
YT is partially supported by World Premier International Research Center Initiative (WPI Initiative), MEXT, Japan.




\bibliographystyle{ytphys}
\small\baselineskip=.97\baselineskip
\bibliography{ref}

\providecommand{\href}[2]{#2}\begingroup\raggedright\begin{thebibliography}{10}

\bibitem{Argyres:1995jj}
P.~C. Argyres and M.~R. Douglas, ``{New Phenomena in $SU(3)$ Supersymmetric
  Gauge Theory},'' \href{http://dx.doi.org/10.1016/0550-3213(95)00281-V}{{\em
  Nucl. Phys.} {\bfseries B448} (1995) 93--126},
\href{http://arxiv.org/abs/hep-th/9505062}{{\ttfamily arXiv:hep-th/9505062}}.

\bibitem{Argyres:1995xn}
P.~C. Argyres, M.~R.~Plesser, N.~Seiberg, and E.~Witten, ``{New
  ${\mathcal{N}}\!=2$ Superconformal Field Theories in Four Dimensions},''
  \href{http://dx.doi.org/10.1016/0550-3213(95)00671-0}{{\em Nucl. Phys.}
  {\bfseries B461} (1996) 71--84},
\href{http://arxiv.org/abs/hep-th/9511154}{{\ttfamily arXiv:hep-th/9511154}}.

\bibitem{Eguchi:1996vu}
T.~Eguchi, K.~Hori, K.~Ito, and S.-K. Yang, ``{Study of ${\mathcal{N}}\!=2$
  Superconformal Field Theories in 4 Dimensions},''
  \href{http://dx.doi.org/10.1016/0550-3213(96)00188-5}{{\em Nucl. Phys.}
  {\bfseries B471} (1996) 430--444},
\href{http://arxiv.org/abs/hep-th/9603002}{{\ttfamily arXiv:hep-th/9603002}}.

\bibitem{Eguchi:1996ds}
T.~Eguchi and K.~Hori, ``{${\mathcal{N}}\!=2$ Superconformal Field Theories in
  Four-Dimensions and A-D-E Classification},'' in {\em The Mathematical Beauty
  of Physics: A Memorial Volume for Claude Itzykson}, J.~Drouffe and J.~Zuber,
  eds., vol.~24 of {\em Advanced Series in Mathematical Physics}, pp.~67--82.
\newblock World Scientific, 1997.
\newblock
\href{http://arxiv.org/abs/hep-th/9607125}{{\ttfamily arXiv:hep-th/9607125
  [hep-th]}}.
\newblock

\bibitem{Gaiotto:2010jf}
D.~Gaiotto, N.~Seiberg, and Y.~Tachikawa, ``{Comments on Scaling Limits of 4D
  ${\mathcal{N}}\!=2$ Theories},''
  \href{http://dx.doi.org/10.1007/JHEP01(2011)078}{{\em JHEP} {\bfseries 1101}
  (2011) 078},
\href{http://arxiv.org/abs/1011.4568}{{\ttfamily arXiv:1011.4568 [hep-th]}}.

\bibitem{Cecotti:2010fi}
S.~Cecotti, A.~Neitzke, and C.~Vafa, ``{R-Twisting and 4D/2D
  Correspondences},''
\href{http://arxiv.org/abs/1006.3435}{{\ttfamily arXiv:1006.3435 [hep-th]}}.

\bibitem{Bonelli:2011aa}
G.~Bonelli, K.~Maruyoshi, and A.~Tanzini, ``{Wild Quiver Gauge Theories},''
  \href{http://dx.doi.org/10.1007/JHEP02(2012)031}{{\em JHEP} {\bfseries 1202}
  (2012) 031},
\href{http://arxiv.org/abs/1112.1691}{{\ttfamily arXiv:1112.1691 [hep-th]}}.

\bibitem{Xie:2012hs}
D.~Xie, ``{General Argyres-Douglas Theory},''
\href{http://arxiv.org/abs/1204.2270}{{\ttfamily arXiv:1204.2270 [hep-th]}}.

\bibitem{Argyres:1996eh}
P.~C. Argyres, M.~R. Plesser, and N.~Seiberg, ``{The Moduli Space of
  ${\mathcal{N}}\!=2$ SUSY {QCD} and Duality in ${\mathcal{N}}\!=1$ SUSY
  {QCD}},'' \href{http://dx.doi.org/10.1016/0550-3213(96)00210-6}{{\em Nucl.
  Phys.} {\bfseries B471} (1996) 159--194},
\href{http://arxiv.org/abs/hep-th/9603042}{{\ttfamily arXiv:hep-th/9603042}}.

\bibitem{Cecotti:2011rv}
S.~Cecotti and C.~Vafa, ``{Classification of complete N=2 supersymmetric
  theories in 4 dimensions},''
\href{http://arxiv.org/abs/1103.5832}{{\ttfamily arXiv:1103.5832 [hep-th]}}.

\bibitem{Alim:2011ae}
M.~Alim, S.~Cecotti, C.~Cordova, S.~Espahbodi, A.~Rastogi, {\em et al.}, ``{BPS
  Quivers and Spectra of Complete ${\mathcal{N}}\!=2$ Quantum Field
  Theories},''
\href{http://arxiv.org/abs/1109.4941}{{\ttfamily arXiv:1109.4941 [hep-th]}}.

\bibitem{Nanopoulos:2010bv}
D.~Nanopoulos and D.~Xie, ``{More Three Dimensional Mirror Pairs},''
  \href{http://dx.doi.org/10.1007/JHEP05(2011)071}{{\em JHEP} {\bfseries 1105}
  (2011) 071},
\href{http://arxiv.org/abs/1011.1911}{{\ttfamily arXiv:1011.1911 [hep-th]}}.

\bibitem{Shapere:1999xr}
A.~D. Shapere and C.~Vafa, ``{BPS Structure of Argyres-Douglas Superconformal
  Theories},''
\href{http://arxiv.org/abs/hep-th/9910182}{{\ttfamily arXiv:hep-th/9910182}}.

\bibitem{Seo:2012ns}
J.~Seo and K.~Dasgupta, ``{Argyres-Douglas Loci, Singularity Structures and
  Wall-Crossings in Pure ${\mathcal{N}}\!=2$ Gauge Theories with Classical
  Gauge Groups},'' {\em JHEP} {\bfseries 1205} (2012) 072,
\href{http://arxiv.org/abs/1203.6357}{{\ttfamily arXiv:1203.6357 [hep-th]}}.

\bibitem{Argyres:1994xh}
P.~C. Argyres and A.~E. Faraggi, ``{The Vacuum Structure and Spectrum of
  ${\mathcal{N}}\!=2$ Supersymmetric $SU(n)$ Gauge Theory},''
  \href{http://dx.doi.org/10.1103/PhysRevLett.74.3931}{{\em Phys. Rev. Lett.}
  {\bfseries 74} (1995) 3931--3934},
\href{http://arxiv.org/abs/hep-th/9411057}{{\ttfamily arXiv:hep-th/9411057}}.

\bibitem{Klemm:1994qs}
A.~Klemm, W.~Lerche, S.~Yankielowicz, and S.~Theisen, ``{Simple Singularities
  and ${\mathcal{N}}\!=2$ Supersymmetric Yang-Mills Theory},''
  \href{http://dx.doi.org/10.1016/0370-2693(94)01516-F}{{\em Phys. Lett.}
  {\bfseries B344} (1995) 169--175},
\href{http://arxiv.org/abs/hep-th/9411048}{{\ttfamily arXiv:hep-th/9411048}}.

\bibitem{Douglas:1996sw}
M.~R. Douglas and G.~W. Moore, ``{D-Branes, Quivers, and ALE Instantons},''
\href{http://arxiv.org/abs/hep-th/9603167}{{\ttfamily arXiv:hep-th/9603167
  [hep-th]}}.

\bibitem{Brandhuber:1995zp}
A.~Brandhuber and K.~Landsteiner, ``{On the Monodromies of ${\mathcal{N}}\!=2$
  Supersymmetric Yang-Mills Theory with Gauge Group SO(2n)},''
  \href{http://dx.doi.org/10.1016/0370-2693(95)00986-U}{{\em Phys. Lett.}
  {\bfseries B358} (1995) 73--80},
\href{http://arxiv.org/abs/hep-th/9507008}{{\ttfamily arXiv:hep-th/9507008}}.

\bibitem{Argyres:1995fw}
P.~C. Argyres and A.~D. Shapere, ``{The Vacuum Structure of ${\mathcal{N}}\!=2$
  Superqcd with Classical Gauge Groups},''
  \href{http://dx.doi.org/10.1016/0550-3213(95)00661-3}{{\em Nucl. Phys.}
  {\bfseries B461} (1996) 437--459},
\href{http://arxiv.org/abs/hep-th/9509175}{{\ttfamily arXiv:hep-th/9509175}}.

\bibitem{Hanany:1995fu}
A.~Hanany, ``{On the Quantum Moduli Space of ${\mathcal{N}}\!=2$ Supersymmetric
  Gauge Theories},'' \href{http://dx.doi.org/10.1016/0550-3213(96)00077-6}{{\em
  Nucl. Phys.} {\bfseries B466} (1996) 85--100},
\href{http://arxiv.org/abs/hep-th/9509176}{{\ttfamily arXiv:hep-th/9509176}}.

\bibitem{Argyres:2007cn}
P.~C. Argyres and N.~Seiberg, ``{S-Duality in ${\mathcal{N}}\!=2$
  Supersymmetric Gauge Theories},''
  \href{http://dx.doi.org/10.1088/1126-6708/2007/12/088}{{\em JHEP} {\bfseries
  0712} (2007) 088},
\href{http://arxiv.org/abs/0711.0054}{{\ttfamily arXiv:0711.0054 [hep-th]}}.

\bibitem{Boalch1}
P.~Boalch, ``{Irregular connections and Kac-Moody root systems},''
  \href{http://arxiv.org/abs/0806.1050}{{\ttfamily arXiv:0806.1050 [math.DG]}}.

\bibitem{Boalch2}
P.~Boalch, ``{Hyperk{\"a}hler manifolds and nonabelian Hodge theory of
  (irregular) curves},'' \href{http://arxiv.org/abs/1203.6607}{{\ttfamily
  arXiv:1203.6607 [math.AG]}}.

\bibitem{Gaiotto:2008ak}
D.~Gaiotto and E.~Witten, ``{S-Duality of Boundary Conditions in
  ${\mathcal{N}}\!=4$ Super Yang-Mills Theory},''
\href{http://arxiv.org/abs/0807.3720}{{\ttfamily arXiv:0807.3720 [hep-th]}}.

\end{thebibliography}\endgroup

\end{document}